\input lecproc.cmm
\def\ref{\noindent\hangindent=1.5cm\hangafter=1}
\def\apj{{ApJ, }}

\def\aa{{A\&A, }}
\def\mn{{MNRAS, }}
\def\nat{{Nature, }}
\def\deg{^{\circ }}

\def\simlt{\lower.5ex\hbox{$\; \buildrel < \over \sim \;$}}
\def\simgt{\lower.5ex\hbox{$\; \buildrel > \over \sim \;$}}
 
\contribution{${\Delta T\over T}$ Beyond Linear Theory}

\contributionrunning{${\Delta T\over T}$ Beyond Linear Theory}

\author{Enrique Mart\'\i nez-Gonz\'alez} 
\address{Instituto de F\'\i sica de Cantabria,
Consejo Superior de Investigaciones Cient\'\i ficas--Universidad de Cantabria, 
Facultad de Ciencias, Avda. Los Castros s/n, 39005 Santander, Spain}

\abstract{The major contribution to the anisotropy of the temperature of the 
Cosmic Microwave Background (CMB) radiation is believed to come from the 
interaction of linear density perturbations with the radiation previous to the
decoupling time. Assuming a standard thermal history for the gas after
recombination, only the gravitational field produced by the linear density
perturbations present on a $\Omega\neq 1$ universe can generate anisotropies
at low z (these anisotropies would manifest on large angular scales).
However, secondary anisotropies are inevitably produced during
the nonlinear evolution of matter at late times even in a universe with a
standard thermal history. Two effects associated to this nonlinear phase can
give rise to new anisotropies: the time-varying gravitational potential of
nonlinear structures (Rees-Sciama RS effect) and the inverse Compton
scattering of the microwave photons with hot electrons in clusters of galaxies
(Sunyaev-Zeldovich SZ effect). These two effects can produce distinct imprints
on the CMB temperature anisotropy. We discuss the amplitude of the anisotropies
expected and the relevant angular scales in different cosmological scenarios.
Future sensitive experiments will be able to probe the CMB anisotropies beyong
the first order primary contribution.} 

\titlea{1}{Introduction}

Cosmic Microwave Background (CMB) temperature anisotropies carry information
on the large scale distribution of matter via the gravitational imprint of
linear density fluctuations on large angular scales ($\simgt
(2\Omega^{1/2})\deg$), the so called Sachs-Wolfe effect for a $\Omega=1$
universe (Sachs and Wolfe 1967) or the generalized Sachs-Wolfe effect for a
$\Omega\neq 1$ universe or in the presence of a cosmological constant $\Lambda$
(Anile and Motta 1967, Wilson 1983, Gouda et al. 1991). In the first case the
gravitational potential is static and the temperature fluctuations are
produced by the potential fluctuations at recombination whereas in the other
cases they are also produced by an integrated effect of the time-varying 
potential along the photon path from recombination to the observer.

On smaller angular
scales the major contribution to the CMB anisotropy comes from photon
fluctuations and the Doppler effect due to the velocity of the last scatterers
at recombination. These primary anisotropies can be erased if the gas in the
universe is reionized at a sufficiently early time ($z\simgt 100$). However, in
this case new secondary anisotropies are generated by the gas motion which
produces a Doppler effect and
by the coupling of velocity and density fluctuations (Vishniac 1987). All
the mentioned effects characterize the shape of the radiation power spectrum
which contains information on the cosmological parameters $\Omega$, $\Lambda$,
the baryonic content $\Omega_b$, the Hubble parameter, the initial matter
density distribution and the thermal history of the universe (see Hu in this
volume for a general review on the CMB anisotropies). 

CMB temperature anisotropies, detected for the first time by Smoot et al.
(1992) with the COBE-DMR experiment, are believed to be 
dominated by the interaction of matter density perturbations and radiation to 
first order in perturbation theory. In the standard scenario these primary 
anisotropies
are produced before neutral hydrogen  formed at recombination  ($z\approx 
1000$) and they are preserved until today. However, secondary anisotropies are
inevitably produced during the nonlinear evolution of matter at late times.   
Even in a flat universe with a standard thermal history (i.e. with no 
reionization of the gas content after recombination), the time-varying
gravitational potential due to the evolving nonlinear structures generates 
new temperature fluctuations (Rees-Sciama RS effect, Rees and Sciama 1968). 
Their amplitude is
estimated to be $\sim 10^{-6}$ and the signatures imprinted in the radiation
power spectrum may distinguish them from the primary anisotropies. We will
discuss the possible detectability of these secondary fluctuations with future
very sensitive experiments.

The Compton scattering of microwave photons by hot electrons in the
intracluster gas also produces new anisotropies at very late times after the
formation of rich clusters of galaxies (Sunyaev and Zeldovich 1972). This
Sunyaev-Zeldovich effect (SZ) produces spectral distortions which translate 
into anisotropies when the angular distribution of temperatures in the sky is 
measured by an experiment at a given frequency band. The peculiar velocity of
clusters also generates anisotropy via Doppler effect but the
temperature amplitude of this
kinematic Sunyaev-Zeldovich effect (KSZ) is expected to be at least an order
of magnitude below the thermal SZ one in the Raileygh-Jeans region of the
spectrum. The SZ effect is several orders of magnitude
greater than the RS effect for a rich cluster. However, since the SZ effect is
restricted to the central regions of rich clusters with a high concentration
of hot gas whereas the RS one is a much more general effect the statistical
average of many structures in the sky may give comparable amplitudes of the
two effects or may even be dominant for the later effect. 

\titlea {2}{The Rees-Sciama Effect} 

Rees and Sciama (1968) first noticed that the time-varying gravitational
potential of evolving nonlinear structures will produce secondary 
anisotropies in the temperature of the CMB. That seminal paper, where an
incorrect Newtonian approach was used to study the anisotropy generated by a
compensated cluster, motivated many works which used a General Relativistic
treatment to model the nonlinear evolution of spherical   
over-densities and under-densities in a uniform background. A simple model
used was that of the "Swiss-cheese" where it is assumed a central spherical 
over-density surrounded by a compensated under-density (Notale 1984, Mart\'\i
nez-Gonz\'alez and Sanz 1990). Also a similar model consisting of a central 
spherical under-dense region compensated by a thin shell was used for cosmic 
voids (Thompson and Vishniac 1987, Dyer and Ip 1988, Mart\'\i nez-Gonz\'alez
and Sanz 1990). A more sophisticated spherical model based on the Tolman
solution of the Einstein equations allows a general density profile (Panek
1992, Arnau et al. 1993, Fang and Wu 1993, Fullana et al. 1996).  

A useful expression to calculate the RS effect produced by non-spherical 
structures and general realistic 
matter distributions can be derived within the potential approximation
(Mart\'\i nez-Gonz\'alez et al. 1990, see below). Chodorowski (1992, 1994)
and Atrio-Barandela and Kashlinski (1992) used that  expression to
calculate the effect of pancake-like collapse of ellipsoids. The results of
all these works for the most prominent structures seen in our local universe
can be summarized as follows. For the large Bootes void of size 
$\approx 60\, h^{-1}$
Mpc located at a distance of $\approx 150\, h^{-1}$ Mpc and with a density of 
galaxies
$\approx 25\%$ of the background density (Kirshner et al. 1981, de Lapparent
et al. 1986), it is estimated  $\Delta T/T\approx a\, few \times 10^{-7}$ with
small variation for different models (Mart\'\i nez-Gonz\'alez and Sanz 1990,
Panek 1992, Arnau et al. 1993). For the Great Attractor located at a distance
of $\approx 60 \, h^{-1}$ Mpc and a density contrast of $\approx 1$ inside a
radius of $\approx 40\, h^{-1}$ Mpc (Lynden-Bell et al. 1987) the estimated
RS effect is $\approx a\, few \times 10^{-6}$ and again the results are fairly
insensitive to the model used (Mart\'\i nez-Gonz\'alez et al. 1990, Panek
1992, Arnau et al. 1994). In the case of the large concentration of galaxies 
known as the Great Wall which forms a flattened structure of dimensions
$\approx 5\times 60\times 170\, h^{-1}$ Mpc at a distance of some $75\, h^{-1}$
Mpc (Geller and Huchra 1989) the effect is also $a \, few \times 10^{-6}$
(Atrio-Barandela and Kashlinski 1992, Chodorowski 1994).

If similar structures are assumed to be at different distances from the
observer in low density universes then it is possible to increase the RS
effect in almost an order of magnitude. For distances $z\sim 1-10$ and
universes with $\Omega\approx 0.2$ great attractor-like structures can produce
an effect of the order of $10^{-5}$ (Arnau et al. 1994) and Bootes-like voids
of $a \, few\times 10^{-6}$ (Fullana et al. 1996), with  
angular scales of a few degrees. Thus, only in open universes the presence
of such large inhomogeneities might leave an observable imprint in the CMB
anisotropy maps.

Below we will describe a useful expression for the RS effect in 
realistic situations and use it to calculate the statistical effect
produced by a distribution of matter. 
   
\titleb {2.1}{The Potential Approximation}

Mart\'\i nez-Gonz\'alez et al. (1990), based on the potential approximation to
general relativity, derived a useful expression for the
temperature fluctuation generated when microwave photons cross a nonlinear
density perturbation $\Delta (t, \vec x)$. If $\varphi(t,\vec x)$ is the
corresponding gravitational potential then the RS effect can be simply written
as 

$$\bigl(\Delta T/T\bigr)_{secondary}=2\int^o_{ls} dt {\partial\varphi\over
\partial t}(t,\vec x)\ \ \ ,
\ \ \nabla^2 \varphi=6\Omega a^{-1}\Delta(t,\vec x)\ \ \ ,\eqno(1)$$
where $a(t)$ is the scale factor normalized to the present time $(a_o=1)$ and 
our units are $c=8\pi G=1$ and the Hubble length at present is
$d_o=2/H_o^{-1}$. Thus, the Rees-Sciama effect is basically the work performed
by a photon travelling through the time-dependent gravitational potential
$\varphi(t,\vec x)$ from recombination to the observer. The previous equation
was initially derived for a flat universe, however, since the scales relevant
for nonlinear structures are always much smaller than the curvature scale it
also applies to open universes. The integral must be performed along the
geodesic associated to the corresponding background 

$$\vec x (t,\vec n )= \lambda (a)\vec n\ \ ,\ \lambda (a)={1-(\Omega
/a+1-\Omega)^{-1/2}\over 1-(1-\Omega)(\Omega /a+1-\Omega)^{-1/2}}\ \ \eqno(2)$$

An order of magnitude estimate of the effect can easily be obtained from
equation (1). For a photon crossing a cluster the anisotropy is

$$ {\Delta T\over T} \sim 2 {\varphi \over c^2} {t_c\over t_d} \sim 2 
{\varphi \over c^2} {v\over c} \sim 2 \Bigl( {v\over c }\Bigr)^3 \ \ \eqno(3)$$
where $\varphi$ is the potential of the cluster, $t_c$ and $t_d$ are the 
crossing time and the dynamical time and $v$ the velocity of collapse. For 
a cluster with $v=1000$ km/s the effect is $\Delta T/T\sim 10^{-7}$. A
statistical distribution of lumps can therefore increase the effect to values
of cosmological interest.

\titleb {2.2}{Second Order Perturbation Theory}

	We will now estimate the RS effect in an open or flat background for a
cold dark matter (CDM) cosmogony. This effect has been recently calculated by
Sanz et al. (1996) within the second order perturbation theory. We will follow
that paper below.

For vanishing pressure, by perturbing the Einstein field equations and
considering perturbations in density up to the 
second order we can obtain the following expression for the density 
fluctuations  

$$\Delta (t,\vec x)=D\delta +D^2\bigl[{5\over 7}\delta^2+\vec\nabla
\delta\cdot\vec\nabla\xi+{2\over 7}\xi_{,ij}\xi^{,ij}\bigr]\ \ \ ,\ \ \
\nabla^2\xi=\delta \eqno(4)$$

\noindent where $\delta (\vec x)\equiv \delta_r (1+z_r)$. Equation (4) is exact
within the 2nd order perturbation theory for a flat background (see Peebles
1980) and it is a good approximation for open universes with $\Omega\simgt
0.1$ as has been shown by Bouchet et al. (1993) and Catelan et al.
(1995). By using equations (2,3) in the expression for the second order effect
(eq. 1), we obtain the following result
 
$$\bigl({\Delta T\over T}(\vec n)\bigr)_{secondary}=\int^o_{\lambda_{ls}} 
d\lambda\, W(\lambda) \phi(\vec x=\lambda\vec n)\ \ ,$$ 
$$W(\lambda)=24\Omega D^2(2f-1){1-(1-\Omega)\lambda\over (1-\lambda)^3}
\eqno(5)$$
where the new potential related to the second order desity perturbation
$\delta_2$ satisfies the following Poisson equation
$$\nabla^2\phi(\vec x)=\delta_2\equiv {5\over 7}\delta^2+\vec\nabla\delta
\cdot\vec\nabla\xi+{2\over 7}\xi_{,ij}\xi^{,ij}\ \ \ .\eqno(6)$$

\noindent In the equation above $D$ is the growing mode of the perturbations
normalized to $1$ at the present time. For a flat universe: $D = a$, whereas
for an open universe (Peebles, 1980):

$$D = {g(x)\over g(x_o)}\ \ \ , x = ({1\over \Omega }-1)a\ \ \ ,
f(x) = {d\ln D\over d\ln a}\ \ \ ,$$

$$g(x) = 1+({3\over x})\Bigl[ 1+(1+{1\over x})^{1/2}\ln \Bigl( (1+x)^{1/2}-
x^{1/2}\Bigr) \Bigr]\ \ \ .\eqno(7)$$

It is common to expand the temperature fluctuations in terms of spherical
harmonics $Y_{lm}$, $\Delta T(\vec n)=\sum_l\sum_{m=-l}^{l}a_{lm}Y_{lm}(\vec 
n)$. The temperature correlation function is given in terms of the radiation
power spectrum as $C(\Theta)=1/(4\pi)\, \sum (2l+1)C_l P_l(\cos(\alpha))$,
with $C_l= <a^2_{lm}>$. In the second order perturbation theory the multipole
component $C_l$ is (we shall not consider the monopole and dipole in the 
calculations of the temperature anisotropy below):

$$C_l={2\over \pi}\int dk k^{-2} P_{(2)}(k)R_l^2(k)\ \ ,
R_l(k)\equiv \int_0^{\lambda_{ls}} d\lambda W(\lambda) j_l(kp)\ .\eqno(8)$$

\noindent Here, $p\equiv \lambda/[1-(1-\Omega)\lambda^2]$, $j_l$ is the Bessel
function of fractional order and
$\lambda_{ls}$ is the distance from the observer to the last scattering 
surface. This equation generalizes the result obtained by Mart\'\i
nez-Gonz\'alez et al. (1992) to open universes.
The function $P_{(2)}(k)$ is the power spectrum associated with the
2nd order density perturbation $\delta_2$
and is related to the power spectrum $P_\xi $ of the time derivative of the
potential $\xi$ by $P_\xi={1\over k^4}P_{(2)}(k)$. The second order
perturbation power spectrum $P_{(2)}(k)$ is given in terms of the first order
power spectrum $P(k)$ by the equation (Goroff et al. 1986, Suto and Sasaki,
1991)

$$P_{(2)}(k)={k^3\over 98(2\pi)^2}\int_0^\infty dr P(kr)\int_{-1}^1 dx
P\bigl(k(r^2+1-2rx)^{1/2}\bigr)$$
$$\Bigl({3r+7x-10rx^2\over r^2+1-2rx}\Bigr)^2\ \ .\eqno(9)$$

\noindent In the limit of small $k$, there is a cancellation of the three
terms contributing
to $\delta_2$, implying that $P_{(2)}(k)$ has very little power on large
scales and in this regime goes like $P_{(2)}(k)\propto k^4$ independently of 
the primordial power spectrum.

Equation (8) for the multipoles $C_l$ can be further simplified by realizing
that the integral to calculate $R_l(k)$ is the product of 
a slowly changing function of $\lambda$, the window $W(\lambda)$, and a
spherical Bessel function (Seljak 1996). Thus, fixing the value of the
argument of $W$ and applying the large $l$ approximation
$\int_o^x j_l(x')=\sqrt(\pi/2l)S(x-l)$, $S$ being the step function, we obtain

$$C_l={1\over l^4} \int_o^{\lambda_{ls}} d\lambda W^2(\lambda)
P_2[l(1-(1-\Omega)\lambda^2)/\lambda]\ \ .\eqno(10)$$

This approximation works very well and in the worst situation for the 
quadrupole moment the error is $\simlt 30\%$ for all $\Omega$ values. 

In relation to a possible secondary contribution coming from 3rd order
density perturbations for open models (in the case of flat models this does not
exist because the 1st order gravitational potential is static), we
have the following comment: the coupling of the 1st order
gravitational potential $\varphi^{(1)}\propto {D\over a}$ with the 3rd order
potential $\varphi^{(3)} \propto {D^3\over a}$ gives a kernel for the
integrated
gravitational effect proportional to ${D^3\over a^2}(f-1)(3f-1)$, whereas
the coupling of the 2nd order gravitational potential $\varphi^{(2)} \propto
{D^2\over a}$ with itself gives a kernel for the integrated
gravitational effect proportional to ${D^3\over a^2}(2f-1)^2$. This second
function is always greater than the first one. Moreover, for quasi-flat models
we expect a negligible contribution from the coupling of 1st-3rd
order perturbations because $f\approx 1$, whereas for low-$\Omega$ models the
integrated gravitational effect due to 2nd-2nd order perturbations is produced
at high-z ($75\%$ of the final effect is produced in the interval $[10,10^3]$
for $\Omega = 0.1$, see next section) where $f\approx 1$, and so there is 
practically no 1st-3rd order contribution. At smaller z some contribution due to
the 1st-3rd coupling is produced but it is
estimated to be always bounded by that
due to the 2nd-2nd coupling at low-z, and this is a small fraction of the
final contribution.

\titleb {2.3} {Results}

The equations derived in the previous section have been applied to calculate
the predicted amplitudes of the multipole components $C_l$ in open universes.
We assume an open or flat CDM model with a Harrison-Zeldovich primordial
spectrum of matter density fluctuations $P(k)=Ak$ and $\Omega_b=0.05$,
$H=50$km s$^{-1}$ Mpc$^{-1}$. In figure 1 we display the multipoles $C_l$ for
$\Omega=1,0.3,0.1$ due to primary anisotropies (upper curves) and due to 2nd
order anisotropies (lower curves). Primary anisotropies are normalized to the
2-year COBE-DMR maps as given by the analysis of Cay\'on et al. (1996) for the
HZ spectrum. Since secondary anisotropies are generated by the relatively
small scale structure, $\simlt 100$Mpc, it is more appropriate to use 
the $\sigma_{16}=1$ normalization, i.e. the rms density fluctuation is unity
at 16 Mpc. The secondary radiation power spectrum peaks at $l\approx 250$ for
all $\Omega$ values, contrary to the linear one where the multipole order,
$l$, of the peak increases for low $\Omega$ values. For the secondary
contribution, the shift of the maximum of the 2nd order power spectrum
$P_{(2)}(k)$ towards large scales (small $k$) for low  $\Omega$ values is 
balanced by the opposite shift of the maximum of the function $R_l(k)$ in
equation (8). 

\begfig 7 cm
\figure{1}{Radiation power spectrum for the primary (upper curves) and
secondary (lower curves) contributions. Solid, dashed and dotted lines
correspond to $\Omega=1,0.3,0.1$, respectively. The secondary contribution is
normalized to $\sigma_{16}=1$, except for the solid upper curve which
represents the flat case with the COBE-DMR normalization.}
\endfig

The generation of the secondary radiation power spectrum with redshift is
shown in figures (2a,b) for $\Omega=1,0.1$ respectively. For $\Omega=1$ more
than $90\%$ of the anisotropy is produced at $z\simlt 10$ (Mart\'\i
nez-Gonz\'alez et al. 1992). In the case of low $\Omega$ models the anisotropy 
is produced at a relatively higher redshift, for $\Omega=0.1$ about $80\%$ of
the effect is generated at a redshift $z<30$. In any case, reionization of the
matter in the universe cannot substantially erase the secondary anisotropy at
such low redshifts.

\begfig 7 cm
\figure{2a}{Generation of the secondary radiation power spectrum with
redshift for a flat universe. Solid, dashed and dotted lines correspond to the
anisotropy generated from redshifts $z=100,10,1$ to the present,
respectively.}
\endfig
\begfig 7 cm
\figure{2b}{The same as figure (2a) but for an open universe with 
$\Omega=0.1$.}
\endfig

Nonlinear scales do not enter in the second-order calculation (eq. 8) 
because for any $\Omega$ value the maximum of $R_l(k)$ is at $k\approx
l\Omega$ and the scales that are contributing to the multipole $l$ have
$k\simlt l$. Thus, for the multipoles $C_l$ up to $l=1000$ only scales $\ge
12$Mpc contribute to the second-order effect. This result agrees with
the recent calculations by Seljak (1996) considering nonlinear scales, up to
$l\sim 1000$ the RS effect is dominated by the second-order calculations.
Tului and Laguna (1995) and Tului et al. (1996) have computed the RS effect by
propagating a bundle of photons through an evolving N-body simulation from the
decoupling to the present time.
However, one must be careful when calculating the nonlinear effect because at 
scales $\simlt 10$Mpc the use of eq. (1) to compute the 
RS effect may not be correct. The reason is that at those small scales higher
orders including spacial derivatives of the gravitational potential,
which have been neglected in the perturbative expansion of the Einstein field
equations to derive eq. (1), may contribute significantly.

The results mentioned above are for the two-point temperature correlation
function where clearly the primary contribution dominates over the secondary
RS one. However, the linear primary effect
gives a null contribution to the three-point correlation since the primordial
matter perturbations generated during the inflationary phase are Gaussian.
Recently, Munshi et al. (1995) and Mollerach et al. (1995) have computed the 
leading contribution to the three-point correlation function which involves the
second-order RS effect. Unfortunately the predicted signal is below the
unavoidable cosmic variance uncertainty.
     
\titlea {3} {The Sunyaev-Zeldovich Effect}
   
Clusters of galaxies not only induce secondary anisotropies in the CMB by
means of gravitational effects. As first noticed by Sunyaev and Zeldovich
(1970,1972) the scattering of microwave photons by hot electrons in the
intracluster gas produces spectral distortions in the blackbody spectrum,
which is known as the thermal SZ effect. Additionally, the peculiar motion of 
the clusters also produces anisotropy in the CMB via Doppler effect, known as
the kinematic SZ effect. In subsections 3.1 and 3.2 we will study in detail
these two effects (see also the recent review by Rephaeli 1995).

The thermal SZ effect has been measured in a number of rich Abell clusters
using three distinct techniques: single-dish radiometry, bolometric
observations and interferometry (see Birkinshaw 1993). Some of the first
convincing detections were those of the clusters A665, A2218 and 0016+16 using
the single-dish OVRO radio telescope at $20$GHz (Birkinshaw et al. 1984). More
recently Herbig et al. (1995) have measured the effect in the nearby Coma 
cluster at a high significant level with the OVRO 5.5m telescope at 32GHz,
deriving a maximum Rayleigh-Jeans decrement of $\Delta T_{RJ}=-505\pm 92\mu$K.
The first detection of the SZ effect in the millimetric region has been
obtained with the bolometric observations of the SUZIE experiment on A2163 at
$2.2$mm, close to the wavelength at which the SZ decrement is maximum 
(Wilbanks et al. 1994). With these bolometric observations it is expected to
measure the positive effect at wavelengths $<1.4$mm (as we will see below
the thermal SZ effect changes sign at that wavelength). The first image of 
the SZ effect was obtained for A2218 with the Ryle interferometer at 15Ghz
(Jones et al. 1993). More recently, images of the rich clusters A773 and
0016+16 have been taken with the Ryle and OVRO arrays (Grainge et al. 1993,
Saunders 1995, Carlstrom et al. 1996) and several other clusters are being
imaged at present (very recently, an image of the SZ decrement towards
A1413 has been taken by Grainge et al. 1996 with the Ryle Telescope). 
Interferometry makes possible to directly compare the SZ 
image with the X-ray image of a cluster and thus study the structure of the 
intracluster gas.

\titleb {3.1} {Thermal SZ Effect}

The inverse Compton scattering of the isotropic CMB radiation ($h\nu <<
m_ec^2$) by a hot ($T_e>>T$), nonrelativistic ($kT_e<<m_ec^2$) Maxwellian 
electron gas is quantitatively described by a simplified version of
the Kompaneets (1957) equation for the rate of change of the photon occupation
number $n$ (Sunyaev and Zeldovich 1980)

$${\partial n \over \partial y}={1\over x^2}{\partial \over \partial x} x^4
{\partial n \over \partial x} \ \ ,$$
$$y=\int_0^\tau {kT_e\over m_e c^2} d\tau\ ,\ x={h\nu\over kT}\ ,\eqno(11)$$
where the nondimensional frequency $x$ has been defined, the optical depth
due to Thompson scattering is $d\tau=n_e\sigma_Tcdt$ with $\sigma_T$ being the
Thomson cross section and $y$ is the Comptonization parameter. Since for the
richest clusters the radiation is only weakly scattered, $y<<1$, and so the
deviations from the Planckian spectrum are small, then it is easy to solve
equation (11) by inserting in the right-hand side of this equation the Planck
function for the occupation number $n_P=1/(e^x-1)$ (Zeldovich and Sunyaev
1969). The solution for the change of spectral intensity $I=2h\nu^3n/c^2$ is

$$\Delta I={2(kT)^3\over (hc)^2}{x^4e^x\over (e^x-1)^2}y \Bigl[x\coth{x\over
2}-4\Bigr]\ \ .\eqno(12)$$

The corresponding change in the thermodinamic temperature is

$$\Delta T=yT[x\coth{x\over 2}-4]\ \ .\eqno(13)$$

For the Raileygh-Jeans region of the spectrum, $x<<1$, equations (12,13) take
the following simple form:

$$\Delta I_{RJ}\simeq -{2(kT)^3\over (hc)^2}x^2y  \ \ , \Delta T_{RJ}\simeq 
-2yT  \ .\eqno(14)$$

Previous solutions are obtained from the Kompaneets equation which is based on
a diffusion approach
and thus is not in principle suitable for Compton scattering in
clusters where the optical depth $\tau<<1$ and consequently most of 
the microwave photons are not scattered even once. Another limitation is the
nonrelativistic treatment used which affects to the calculations for the 
richest clusters with temperatures $kT_e\sim 10keV$ and produces significant
errors in the Wien region (Fabbri 1981, Raphaeli 1995).
Even so, except for the extreme cases in the distribution of clusters, 
solutions (12,13) provide an adequate description of the SZ effect in the 
whole spectral range of interest (for the richest clusters the solutions are 
still accurate in the Raileygh-Jeans region). 

An interesting property of equations (12,13) is that the spectral change
is zero at $\lambda\approx 1.4$mm ($\nu\approx220$GHz) being negative and
positive above and below that wavelength, respectively. This characteristic 
change in the sign of the SZ effect is
still waiting to be observed (a tentative detection of the sign change has
been recently claimed by Andreani et al. 1996 for the clusters A2744 and
S1077 observed at 1.2 and 2 mm with the 15m SEST antenna in Chile).  
 
We can now estimate the amplitude of the effect for a cluster of galaxies.
Considering the nearby, well-studied Coma cluster, we know from X-ray
data that the properties of the cluster atmosphere consist of a central 
density of
$n_e\approx  2.9\times 10^{-3}$cm$^{-3}$, a radius $r_c\approx 10'.5$
($\approx 0.43$Mpc for $H=50$Km s$^{-1}$ Mpc$^{-1}$) and a temperature
$T_e\approx 9.1$keV (Hughes et al. 1988, Briel et al. 1992). From these values
we can easily derive an approximate optical depth of $\tau\approx n_e\sigma_T
2r_c\approx 5.1\times 10^{-3}$ and a
maximum decrement in the Raileygh-Jeans region of $\Delta T\approx 500\mu$K,
very close to the value derived from the observations with the OVRO 5.5m
telescope (Herbig et al. 1995). 

Combining the SZ measurement
with the parameters of the model  atmosphere of the cluster derived from the 
X-ray data one
can determine the angular diameter distance to a cluster and thus deduce  
the Hubble constant H$_o$ without relying on any cosmic distance ladder. 
Estimates of H$_o$ have already been made for
several clusters of galaxies providing an accuracy for H$_o$ similar to other
methods (see for instance the recent estimate from the SZ measurement in 
Coma by Herbig et al. 1995). The major uncertainty of this method originates
from systematic effects related to the X-ray model, departures
from spherical symmetry and possible existence of clumpiness in the gas
distribution (Birkinshaw et al. 1991).

\titleb {3.2} {Kinematic SZ Effect}

The peculiar velocity of a cluster also modifies the  intensity of the CMB
via the Doppler effect. Because of its nature, this kinematic SZ effect 
produces
fluctuations in the CMB temperature which are independent of frequency, in
contrast to the thermal SZ. The change in temperature is therefore given by

$$\Delta T_k = -\tau T{v_r\over c} \eqno(15)$$
where $v_r$ is the radial velocity of the cluster with positive sign for
recession. The change in the
intensity $\Delta I_k$ can be easily obtained  from the temperature
fluctuation using the Planck spectrum 

$$\Delta I_k=-{2(kT)^3\over (hc)^2}{x^4e^x\over (e^x-1)^2}\tau {v_r\over c}\ . 
\eqno(16)$$

For a rich cluster like Coma and assuming a peculiar velocity of $500$km 
s$^{-1}$ the
expected change in the temperature is only of $\Delta T_k\approx 20 \mu$K,
much smaller than the thermal effect. It is interesting to notice that the
peculiar velocity of a cluster can be determined by measuring its thermal and
kinematic effects and knowing the spatial distribution of the gas temperature
from X-ray spectral data. Future very sensitive multifrequency ground-based,
balloon and satellite 
experiments surveying large areas of the sky  at a resolution of $\simlt 10$
arcmin (see e.g. the proposed satellite COBRAS/SAMBA to ESA M3, Presentation
of Assesment Study Results 1994) will be able 
to measure the
thermal and kinematic SZ effects  in $\sim 1000$ clusters. This information
combined with X-ray spectral data of the same clusters will allow a reliable 
measurement of the bulk motion in a volume of $\sim 1$ Gpc (Haehnelt and
Tegmark 1995). 

\titleb {3.3} {Contribution to the CMB fluctuations}

In the previous subsections we have studied the SZ effect expected when the
microwave photons cross the intracluster gas of hot electrons in a rich
cluster of galaxies. Below we will discuss the contribution of the 
emsemble of clusters to the CMB temperature maps.  

Considerable work has been done to estimate the rms temperature 
fluctuations originated by the SZ effect. The usual approach is to consider
an evolution for the cluster mass function that corresponds to a previously
assumed theoretical model of density perturbations (Cole and Kaiser 1989;
Markevitch et al. 1991, 1992; Makino and Suto 1993; Bartlett and Silk 1994;
Colafrancesco et al. 1994; De Luca et al. 1995). The cluster mass function can
be represented by the Press-Schechter formula (Press and Schechter 1974) 
which gives the comoving number density of collapsed objects per mass interval
at a given redshift. 
A collapsed object is formed when its density contrast
reaches the value $\delta_c=1.68$ as given by the spherical collapse model. 
Assuming that
the cluster is virialized to a given virial radius $R_v$ the
temperature of the gas can be related to the mass of the cluster. The electron
density profile is usually parametrized with a $\beta$-model with the exponent
$\beta\approx 0.75$ from observations. 
Typical rms
values of the temperature fluctuations obtained with this formalism are of 
the order $\Delta T/T\sim 10^{-6}$.

A different approach has been considered by Ceballos and Barcons (1994) who 
used an empirically based model for the mass function of the intracluster 
mass. They use a
parametrization of the X-ray luminosity function and its negative evolution 
(Edge et al. 1990) which supports the idea that the number of luminous 
clusters decreases with $z$ (Giogia et al. 1990; Henry et al. 1992) and
therefore the amount of hot gas available for the inverse Compton scattering
of the CMB photons is consequently limitted. The
result is a negligible SZ contribution to the rms temperature fluctuations of
$\Delta T/T\simlt 10^{-7}$. The contribution is dominated by the hottest
clusters which are easily identified and thus its SZ signal can be removed
from the temperature map. 

The evidence for fewer X-ray luminous clusters at $z>0.2$ than locally shown 
by the Einstein Extended Medium Sensitivity Survey (EMSS) should be confirmed 
using a
complete, X-ray selected sample. The present ongoing cluster surveys based on 
ROSAT data will measure the X-ray luminosity function at redshifts $>0.2$ and 
thus will determine the amount of hot gas at higher redshifts. Evidence of 
the marked decline in the volume density of luminous clusters with redshift 
has been recently presented with the X-ray RIXOS survey up to $z\approx 0.6$ 
by Castander et al. (1996)  (this claim is, however, based on a sample of only 
13 clusters).   

\titlea {4} {Summary}

The RS and SZ effects, associated with the late, nonlinear phase of structure
formation, can give rise to secondary anisotropies.  The RS effect is produced
by the time-varying gravitational potential of the evolving matter density
perturbations whereas the SZ one is due to the inverse Compton scattering of
microwave photons by hot electrons in the intracluster gas.
These two effects produce
distinct signatures on the CMB temperature anisotropy: contrary to the
RS effect the SZ one is frequency dependent. 

The secondary anisotropy
generated by the SZ effect is probably dominated by the more luminous 
clusters and is generated at low redshifts $z\simlt 1$, if the negative 
evolution in the X-ray luminosity function of clusters
with redshift shown by the EMSS
is confirmed. The amplitude of the
signal is very sensitive to the amount of hot gas available at moderate and
high reshift, $z\simgt 0.5$. Thus, predictions based on theoretical models 
like 
the CDM one, which imply an increase in the amplitude of the X-ray luminosity
function with redshift, give anisotropies $(\Delta T/T)_{SZ} \sim 10^{-6}$ 
whereas
calculations which take into account the observed decline of luminous clusters
with $z$ suggest much smaller values.

The amplitude of the RS effect for realistic models of structure formation is 
$(\Delta T/T)_{RS}\simlt 10^{-6}$, being higher for universes with
high $\Omega$. The signal is very sensitive to the normalization of the matter
power spectrum being directly proportional to its amplitude. Early
reionization of the matter in the universe, which can strongly affect the
primary anisotropies, would not appreciably change the secondary ones.
Future very sensitive satellite experiments might be able to test the
gravitational instability theory beyond the linear theory.

\bigskip
\noindent{\sl Acknowledgements:} The author would like to thank Jose Luis
Sanz and Sergio Torres for their comments on the manuscript. 

\begrefchapter{References}

\ref Andreani, P. et al. 1996, \apj  in press

\ref Anile, A.M. and Motta, S. 1967, \apj  207, 685

\ref Arnau, J.V., Fullana, M.J., Monreal, L. and S\'aez, D. 1993, \apj  
402, 359

\ref Arnau, J.V., Fullana, M.J. and Saez, D. 1994, \mn 268, L17 

\ref Atrio-Barandela, F. and Kashlinski, A. 1992, \apj 390, 322

\ref Bartlett, J. and Silk, J. 1994, \apj  407, L45

\ref Birkinshaw, M., Gull, S.F., Hardebeck, H.E. 1984, \nat  309, 34

\ref Birkinshaw, M., Hughes, J.P. and Arnaud, K.A. 1991, \apj 379, 466

\ref Birkinshaw, M. 1993, in Proc. Present and Future of the Cosmic Microwave
Background, eds. J.L. Sanz, E., Mart\'\i nez-Gonz\'alez and L. Cay\'on
(Springer-Verlag)

\ref Blumenthal, G.R., Nicolaci Da Costa, L., Goldwirth, D.S., Lecar,
M. and Piran, T. 1992, \apj 388, 234

\ref Briel, U.G., Henry, J.P. and B\"ohringer, H. 1992, \aa 259, L31 
      
\ref Bond, J.R., Carr, B. and Hogan, C.J. 1991, \apj  367, 420

\ref Bouchet, F., Juszkievicz, R., Colombi, S. and Pellat, R. 1993, preprint

\ref Carlstrom, J.E., Joy, M. and Grego, L. 1995, \apj 456, L75

\ref Castander, F.J. et al. 1996, \nat  in press

\ref Catelan, P., Lucchin, F., Matarrese, S. and Moscardini, L. 1995, preprint

\ref Cay\'on, L. 1995, in this volume

\ref Cay\'on, L., Mart\'\i nez-Gonz\'alez, E., Sanz, J. L., Sugiyama, N.
and Torres, S. 1996, MNRAS, in press

\ref Ceballos, M.T. and Barcons, X. 1994, \mn 271, 817
                
\ref Chodorowski, M. 1992, \mn 259, 218 

\ref Chodorowski, M. 1994, \mn 266, 897

\ref Colafrancesco, S., Mazzotta, P., Rephaeli, Y. and Vittorio, N. 1994, \apj
433, 454

\ref Cole, S. and Kaiser, N. 1988, \mn 233, 637

\ref Dyer, C.C. and Ip, P.S.S. 1988, \mn 235, 895

\ref Edge, A.C., Stwart, G.C., Fabian, A.C. and Arnaud, K.A. 1990, \mn 245,559

\ref Fabbri, R. 1981, Astrophys. Space Sci., 77, 529

\ref Fang, L. and Wu,  X. 1993, \apj , 408, 25

\ref Fullana, M.J., Arnau, J.V. and S\'aez, D. 1996, \mn in press
   
\ref Geller, M.J. and Huchra, J. 1989, Science, 246, 897

\ref Giogia, I.M., Henry, J.P., Maccacaro, T., Morris, S.L., Stocke, J.T. and
Wolter, A. 1990, \apj 356, L35

\ref Goroff, H., Gristein, B., Rey, S.-J. and Wise, M.B. 1986, \apj 311, 6

\ref Gorski, K. M., Ratra, B., Sugiyama, N. and Banday, A. J., 1995, \apj
444, L65.

\ref Gouda, N., Sugiyama, N. and Sasaki, N. 1991, Prog. Theor. Phys.,
85, 1023

\ref Grainge, K., Jones, M., Pooley, G., Saunders, R. and Edge, A. 1993, \mn
265, L57

\ref Grainge, K., Jones, M., Pooley, G., Saunders, R., Baker, J., Haynes, T.
and Edge, A. 1996, \mn 278, L17

\ref Haelhnelt, M.G. and Tegmark, M. 1995, preprint

\ref Henry, J.P., Giogia, I.M., Maccacaro, T., Morris, S.L., Stocke, J.T. and
Wolter, A. 1992, \apj 386, 408

\ref Herbig, T., Lawrence, C.R. and Readhead, A.C.S. 1995, \apj 449, L5

\ref Hu, W. 1995, in this volume

\ref Hughes, J.P., Gorestein, P. and Fabricant, A. 1988, \apj 329, 82

\ref Jones, M., Saunders, R., Alexander, P., Birkinshaw, M. and Dillon, N. 
1993, \nat 365, 320

\ref Kirshner, R.P., Oemler, A., Schechter, P.L. and Schectman
, S.A. 1981, \apj 248, L57               

\ref Kompaneets, A.S. 1957, Sov. Phys. JETP, 4,730

\ref de Lapparent, V., Geller, M.J. and Huchra, J. 1986, \apj 302, L1

\ref De Luca, A., D\'esert, F.X. and Puget, J.L. 1995, \aa 300, 335

\ref Lynden-Bell, D., Faber, S.M., Burstein, D., Davies, R.L.
, Dressler, A., Terlevich, R. and Wegner, G. 1988, \apj 326, 19

\ref Makino, N. and Suto, Y. 1993, \apj 405, 1

\ref Markevitch, M., Blumenthal, G.R., Forman, W., Jones, C. and Sunyaev, R.A.
1991, \apj 378, L33

\ref Markevitch, M., Blumenthal, G.R., Forman, W., Jones, C. and Sunyaev, R.A.
1992, \apj 395, 326

\ref Mart\'\i nez-Gonz\'alez, E. and Sanz, J.L. 1990, MNRAS, 247, 
473

\ref Mart\'\i nez-Gonz\'alez, E. and Sanz, J.L. 1995, Astro. Lett. and
Comm., 32, 89

\ref Mart\'\i nez-Gonz\'alez, E., Sanz, J.L. and Silk, J. 1990, \apj , 355, L5
            
\ref Mart\'\i nez-Gonz\'alez, E., Sanz, J.L. and Silk, J. 1992, Phys. Rev. D, 
46, 4193

\ref Mart\'\i nez-Gonz\'alez, E., Sanz, J.L. and Silk, J. 1994, \apj , 436, 1

\ref Meszaros, A. 1994, \apj , 423, 19
   
\ref Mollerach, S., Gangui, A., Lucchin, F. and Matarrese, S. 1995,
\apj 453, 1

\ref Munshi, D., Souradeep, T. and Starobinsky, A.A. 1995, \apj  454, 552

\ref Nottale, L. 1984, \mn 206, 713

\ref Panek, M. 1992, \apj, 388, 225

\ref Peebles, P. J. E. 1980, The Large Scale Structure of the
Universe, (Princeton, Princeton University Press, 1980)
       
\ref Press, W.H. and Schechter, P. 1974, \apj 187, 425

\ref Raphaeli, Y. 1995, \apj 445, 33

\ref Raphaeli, Y. 1995, Annu. Rev. Astron. Astrophys., 33, 541

\ref Ratra, B. and Peebles, P. J. E., 1994, \apj, 432, L5.
   
\ref Rees, M.J. and Sciama, D.W. 1968, \nat 217, 355

\ref Sachs, R.K. and Wolfe, A.N. 1967, \apj , 147, 73

\ref Saez, D., Arnau, J.V. and Fullana, M.J. 1993, MNRAS, 263, 
681

\ref Sanz, J.L. and Cay\'on, L. 1996, Proc. Conf. Mapping, Measuring and
Modelling the Universe, eds. P. Coles, V. Mart\'\i nez and M.J. Ponz (ASP
Conf. Series)
       
\ref Sanz, J.L., Mart\'\i nez-Gonz\'alez, E., Cay\'on, L., Silk, J. 
and Sugiyama, N. 1995, \apj , in press.

\ref Seljak, U. 1996, \apj , in press.

\ref Smoot, G. et al. 1992, \apj 396, L1

\ref Saunders, R. 1995, Astrophys. Lett. Commun., 32, 339
  
\ref Sunyaev, R.A. and Zeldovich, Y.B. 1970, Astrophys. Space Sci., 7, 3

\ref Sunyaev, R.A. and Zeldovich, Y.B. 1972, Comments Astrophys. Space 
Phys., 4, 173

\ref Sunyaev, R.A. and Zeldovich, Y.B. 1980, Annu. Rev. Astron.
Astrophys., 18, 537

\ref Suto, Y. and Sasaki, M. 1991, Phys. Rev. D, 66, 265

\ref Thompson, K.L. and Vishniac, E.T. 1987, \apj 313, 517

\ref Tului, R. and Laguna, P. 1995, \apj 445, L73

\ref Tului, R., Laguna, P. and Anninos, P. 1996, preprint
            
\ref Vishniac, E.T. 1987, \apj , 322, 597

\ref Wilbanks, T.M., Ade, P.A.R., Fisher, M.L., Holzapfel, W.L. and Lange,
A.E. 1994, \apj 427, 75

\ref Wilson, M.L. 1983, \apj 420, 1

\ref Zeldovich, Y.B. and Sunyaev, R.A. 1969, Astrophys. Space Sci., 4,
301

\endref

\byebye
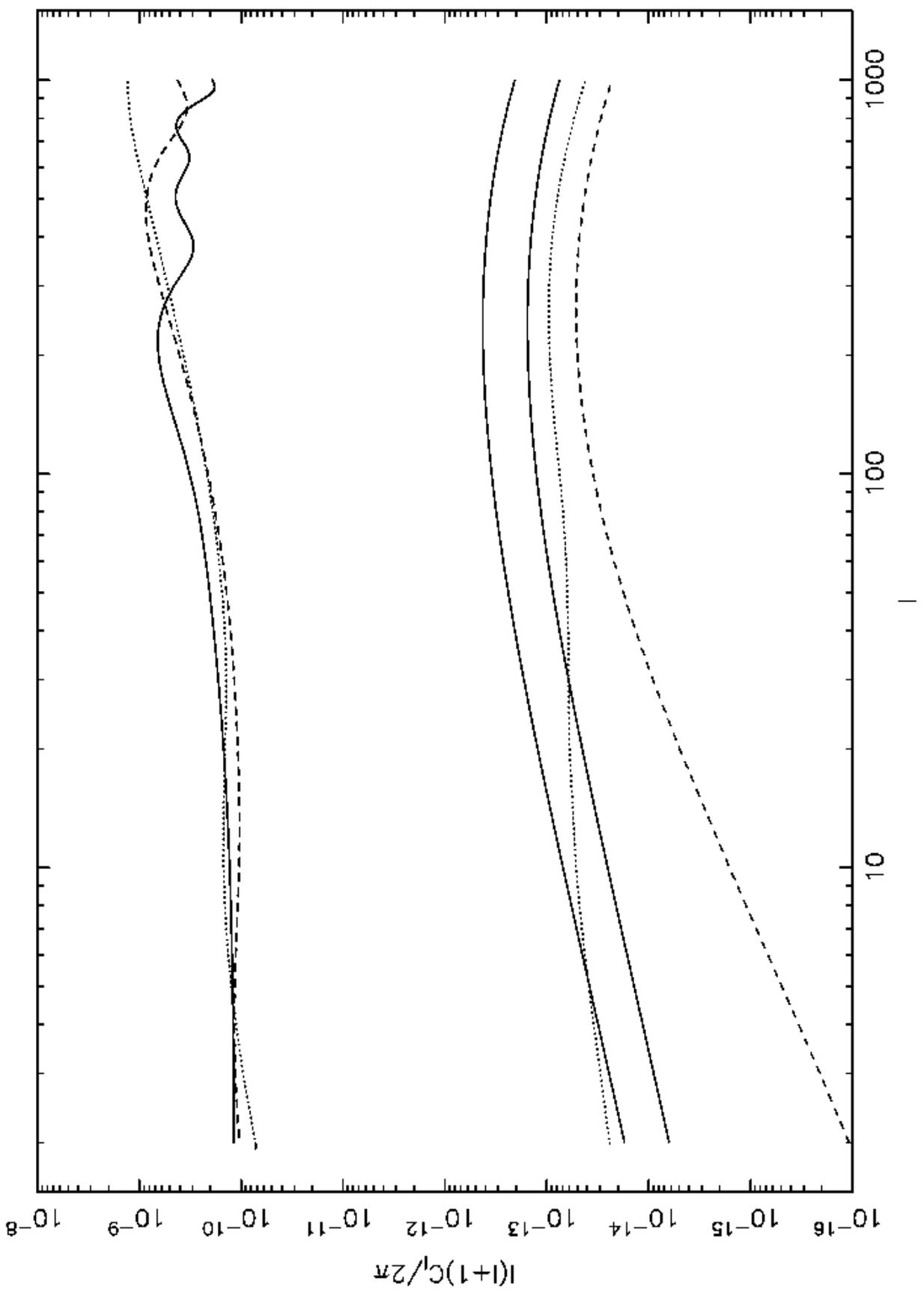
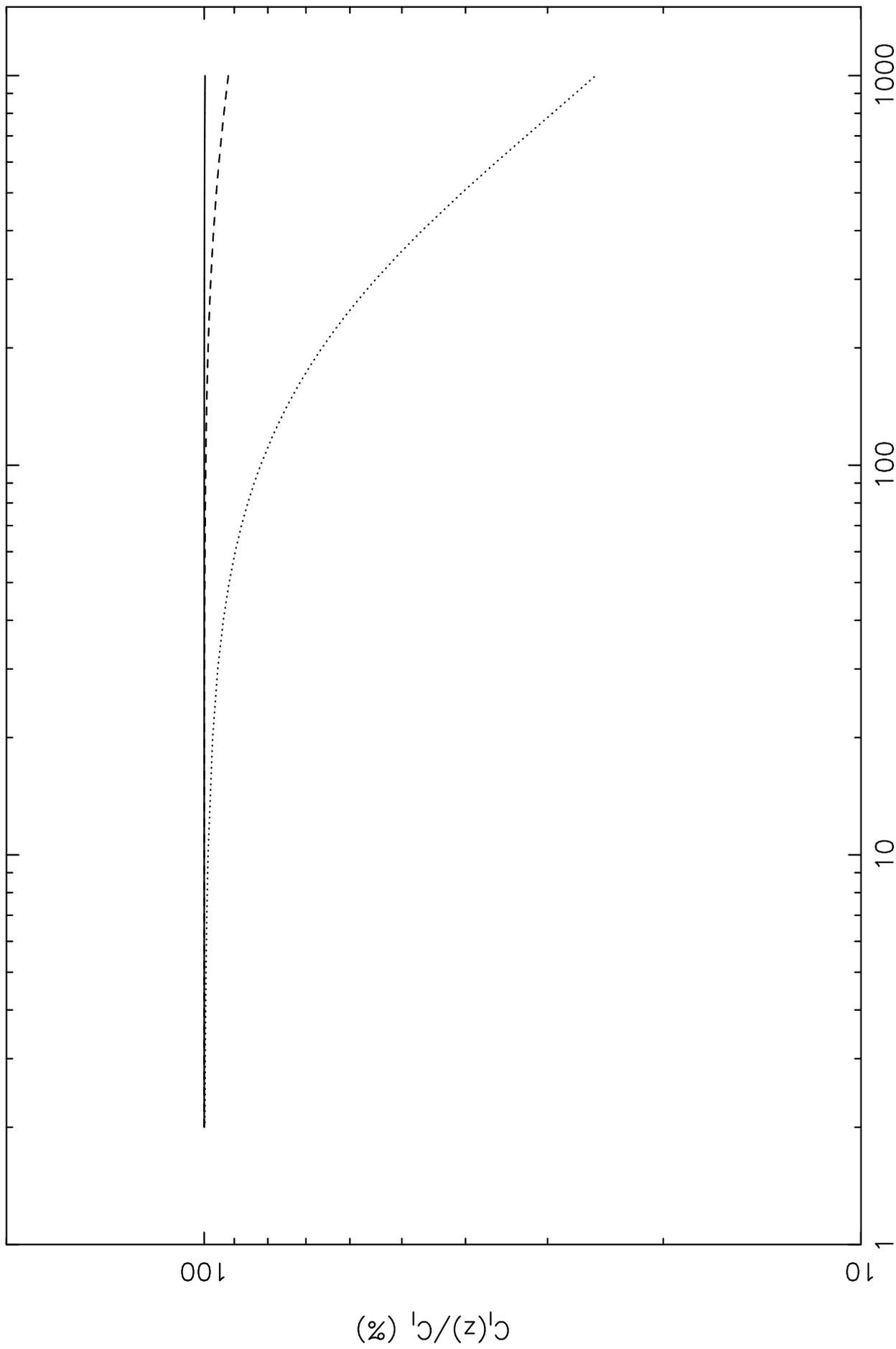
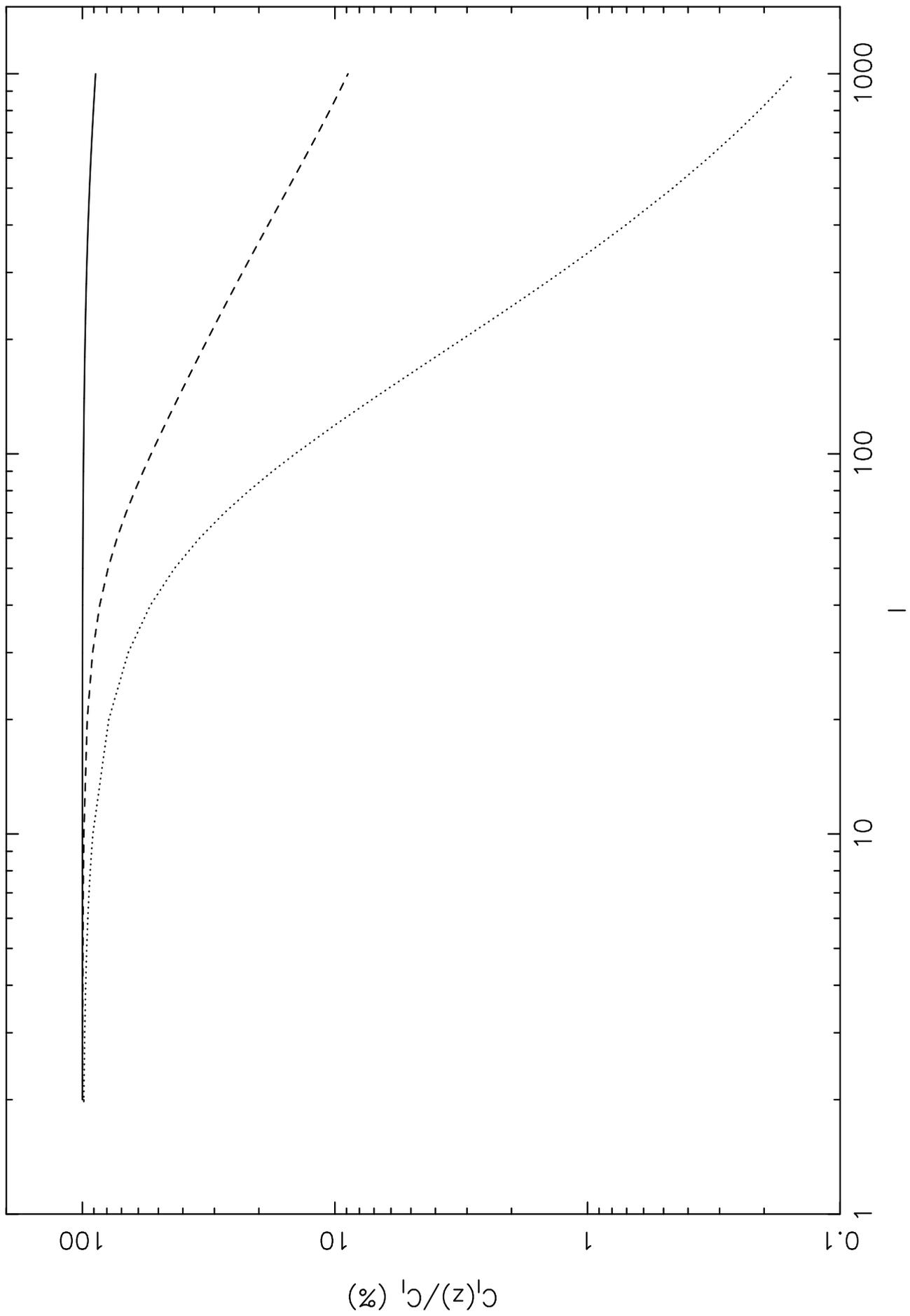
